\documentclass[12pt]{article}
\usepackage{epsfig}
\newcommand{\bq}{\begin{eqnarray}}
\newcommand{\eq}{\end{eqnarray}}
\usepackage{subcaption}

\begin{document}

\title{Swaption Prices in HJM model.\\
Nonparametric fit.}
\author{V.M. Belyaev\\
\it Allianz Life,
Minneapolis, USA}

\maketitle

\begin{abstract}
Closed form formulas for swaption prices in HJM model are derived.
These formulas are used for nonparametric fit of deterministic forward volatility.
It is demonstrated that this formula and non-parametric fit works very well and can be used to identify arbitrage opportunities.
\end{abstract}
\section{Introduction}

In this paper, we propose a closed form formula for swaption prices in HJM model \cite{HW},\cite{HJM} in case of single and dual yield curves.
Here we assume that swaption prices can be calculated in the limit of small forward volatilities.
It is verified that obtained formula works very well and can be used to calibrate all existing ATM swaptions.
Accuracy of this formula is better than Rebonato's formula\cite{Reb} in case of BGM Libor Model\cite{BGM}.
Note, that the formula is applicable in case of multi factor model.

Suggested calibration procedure can be useful for pricing swaptions and  to identify arbitrage opportunities.

\section{Closed Form Formula}
Swap contract value at time $t=0$ can be presented in the following form
\bq
val(t=0)=B(0,T)-B(0,T_N)-\frac{r_X}\nu\sum_{n=1}^NB(0,T_n);
\label{val}
\eq
where $r_X$  swap contract rate; $\nu$ is a frequency of payments; $T_n$ times of $N$-payments; $T$ is a contract start time;
\bq
B(t,T)=e^{-\int_t^Tf(t,\tau)d\tau};
\eq
is a Z-bond price at time $t$, $f(t,\tau)$ is a forward rate.

The initial swap value is equal to zero if
\bq
r_X=r_S=\frac{B(0,T)-B(0,T_N)}{\frac1\nu\sum_{n=1}^NB(0,T_n)};
\eq
where $r_S$ is a swap rate.

Below we consider two options to calculate swap and swaption values.

\subsection{Single Curve}
Heath, Jarrow, and Morton\cite{HJM} forward rate dynamics is:
\bq
df(t,T)=\alpha(t,T)dt+\sigma(t,T)dW(t);
\label{hjm}
\eq
where $\sigma(t,T)$ is a deterministic forward rate volatility; $dW(t)$ is Brownian motion; and
\bq
\alpha(t,T)=\sigma(t,T)\int_t^T\sigma(t,\tau)d\tau;
\label{drift}
\eq
is a drift. This drift can not be arbitrary chosen but it depends on volatility to satisfy arbitrage free conditions for bonds.

Distribution of discounted  swap contract values at time $T$ is
\bq
PV(val(T))=e^{-\int_0^Tr(t)dt}\left(
B(T,T)-B(T,T_N)-\frac{r_X}\nu\sum_{n=1}^NB(T,T_n)
\right);
\eq
where $r(t)=f(t,T)$ is a short term interest rate.

Using the following formula for distribution of discounted bond values in HJM model
\bq
e^{-\int_0^tr(\tau)d\tau}B(t,T)=B(0,T)e^{-\int_0^td\tau\int_\tau^T\alpha(\tau,t_1)dt_1-\int_0^tdW(\tau)\int_\tau^T\sigma(\tau,t_1)dt_1};
\label{distr}
\eq
we obtain  distribution of discounted swap values at time $T$:
\bq
e^{-\int_0^Tr(\tau)d\tau}\left(
B(T,T)-B(T,T_N)-\frac{r_X}\nu\sum_{n=1}^NB(T,T_n)
\right)=
\nonumber
\\
=B(0,T)e^{-\int_0^Td\tau\int_\tau^T\alpha(\tau,t)dt-\int_0^TdW(\tau)\int_\tau^T\sigma(\tau,t)dt}-
\nonumber
\\
-B(0,T_N)e^{-\int_0^Td\tau\int_\tau^{T_N}\alpha(\tau,t)dt-\int_0^TdW(\tau)\int_\tau^{T_N}\sigma(\tau,t)dt}-
\nonumber
\\
-\frac{r_X}\nu \sum_{n=1}^NB(0,T_n)e^{-\int_0^Td\tau\int_\tau^{T_n}\alpha(\tau,t)dt-\int_0^TdW(\tau)\int_\tau^{T_n}\sigma(\tau,t)dt}.
\label{distr2}
\eq

Note that distribution (\ref{distr}) satisfied the following condition
\bq
B(0,T)=\left<
e^{-\int_0^tr(\tau)d\tau}B(t,T)
\right>;
\eq
for any $t$ when drift term $\alpha(t,T)$ is determined according to Eq.(\ref{drift}).

In the limit of small volatility $\sigma\to 0$ the leading term of distribution (\ref{distr2}) is
\bq
\frac{r_S-r_X}\nu\sum_{n=1}^NB(0,T_n)+
\nonumber
\\
+\frac{r_X}\nu\sum_{n=1}^NB(0,T_n)
\int_0^TdW(\tau)\int_\tau^{T_n}\sigma(\tau,t)dt+
\nonumber
\\
+B(0,T_N)\int_0^TdW(\tau)\int_\tau^{T_N}\sigma(\tau,t)dt-
\nonumber
\\
-B(0,T)\int_0^TdW(\tau)\int_\tau^{T}\sigma(\tau,t)dt.
\eq
So, distribution of discounted swap contract values at time $T$ is:
\bq
d=\frac{r_S-r_X}\nu\sum_{n=1}^NB(0,T_n)+\Sigma(T,N)\xi\sqrt{T}+o(\sigma\sqrt{dt});
\label{distr0}
\eq
where
\bq
& & \Sigma^2(T,N)=\frac1T\int_0^Tv^2(t,N)dt;
\label{form}
\nonumber
\\
& & v(t,N)=B(0,T)\int_t^T\sigma(t,\tau)d\tau-B(0,T_N)\int_t^{T_N}\sigma(t,\tau)d\tau-
\nonumber
\\
& & -\frac{r_X}\nu\sum_{n=1}^NB(0,T_n)
\int_t^{T_n}\sigma(t,\tau)d\tau.
\eq
ATM swaption value, when $r_X=r_S$ is
\bq
\int [d]_+e^{-\frac12\xi^2}\frac{d\xi}{\sqrt{2\pi}}=\Sigma(T)\sqrt{\frac{T}{2\pi}}.
\label{ATM}
\eq
where 
\bq
[f(x)]_+=\left\{\begin{array}{rl}
0; & if\; f(x)\leq 0 \\
f(x); & if \; f(x)>0
\end{array}\right.
.
\eq

Swap values can be calculated using day count for float legs. It can be
approximated by factor A = 365.25/360\footnote{
Author thanks E. Tsiper for this remark.
}. In this case, ATM swaption value
can be calculated from eq.(13) with the following change:
\bq
& & v(t,N)=A\left(B(0,T)\int_t^T\sigma(t,\tau)d\tau-B(0,T_N)\int_t^{T_N}\sigma(t,\tau)d\tau\right)-
\nonumber
\\
& & -\frac{r_X}\nu\sum_{n=1}^NB(0,T_n)
\int_t^{T_n}\sigma(t,\tau)d\tau.
\eq

Note, that this approach can be completed in case of multi-factor model as well. 
%For example, in case of noncorrelated factors it will leads to the following replacement:
%\bq
%\sigma^2(t,\tau)=\sum_{n=1}^N\sigma^2_n(t,\tau);
%\eq
%where $\sigma^2_n(t,\tau)$ is $n$-th factor volatility.

\subsection{OIS discounting}
In case of OIS discounting we can assume that Libor-OIS spread is a deterministic function.
Then ATM swaption value is
\bq
PV(val(T))=e^{-\int_0^Tr(t)dt}\left(A
\sum_{n=1}^N\left(
\frac{B(T,T_{n-1})S(T_{n-1})}{S(T_n)}-B(T,T_n)
\right)-
\right.
\nonumber
\\
\left.
-\frac{r_X}\nu\sum_{n=1}^NB(T,T_n)
\right);
\label{OisSw}
\eq
where $S(t)=e^{-\int_0^ts(\tau)d\tau}$ is deterministic Libor-OIS spread.

From (\ref{OisSw}) the ATM strike is
\bq
r_{ATM}=A\nu\frac{\sum_{n=1}^NB(0,T_{n-1})\frac{S(T_{n-1})}{S(T_n)}-B(0,T_n)}{
\sum_{n=1}^NB(0,T_n)
\label{RATM}
};
\eq
and initial swaption value is
\bq
e^{-\int_0^Tr(t)dt}\left(A
\sum_{n=1}^N\left(
\frac{B(T,T_{n-1})S(T_{n-1})}{S(T_n)}-B(T,T_n)
\right)-
\right.
\nonumber
\\
\left.
-\frac{r_X}\nu\sum_{n=1}^NB(T,T_n)
\right)=
\nonumber
\\
=A\sum_{n=1}^N\left(\frac{S(T_{n-1})}{S(T_n)}B(0,T_{n-1})
e^{-\int_0^Td\tau\int_\tau^{T_{n-1}}\alpha(\tau,t)dt-\int_0^TdW(\tau)\int_\tau^{T_{n-1}}\sigma(\tau,t)dt}-
\right.
\nonumber
\\
\left.
-B(0,T_n)e^{-\int_0^Td\tau\int_\tau^{T_{n}}\alpha(\tau,t)dt-\int_0^TdW(\tau)\int_\tau^{T_{n}}\sigma(\tau,t)dt}
\right)-
\nonumber
\\
-\frac{r_X}{\nu}\sum_{n=1}^NB(0,T_n))e^{-\int_0^Td\tau\int_\tau^{T_{n}}\alpha(\tau,t)dt-\int_0^TdW(\tau)\int_\tau^{T_{n}}\sigma(\tau,t)dt}.
\eq
So, in the limit $\sigma\to 0$ for ATM swaption we have
\bq
d(T,N)=\Sigma(T)\xi\sqrt{T}+o(\sigma\sqrt{dt});
\label{distr1}
\eq
where
\bq
& & \Sigma^2(T,N)=\frac1T\int_0^Tv^2(t,N)dt;
\nonumber
\\
& & v(t,N)=\left(A+\frac{r_{ATM}}{\nu}\right)\sum_{n=1}^NB(0,T_n)\int_t^{T_n}\sigma(t,\tau)d\tau
-
\nonumber
\\
& & -\sum_{n=1}^NB(0,T_{n-1})\frac{S(T_{n-1})}{S(T_n)}
\int_t^{T_{n-1}}\sigma(t,\tau)d\tau.
\label{v}
\eq
Discretized version of (\ref{v}) has the following form:
\bq
v(t_i,N)=\left(A+\frac{r_{ATM}}{\nu}\right)\sum_{n=1}^NB(0,T_n)\sum_{j=i}^{n-1}\sigma(t_i,t_j)dt-
\nonumber
\\
-A\sum_{n=1}^NB(0,T_{n-1})\frac{S(T_{n-1})}{S(T_n)})\sum_{j=i}^{n-2}\sigma(t_i,t_j)dt;
\label{sf}
\eq
where $i=-T/dt$.

\begin{figure}[h]
\begin{center}
\includegraphics[width=100mm]{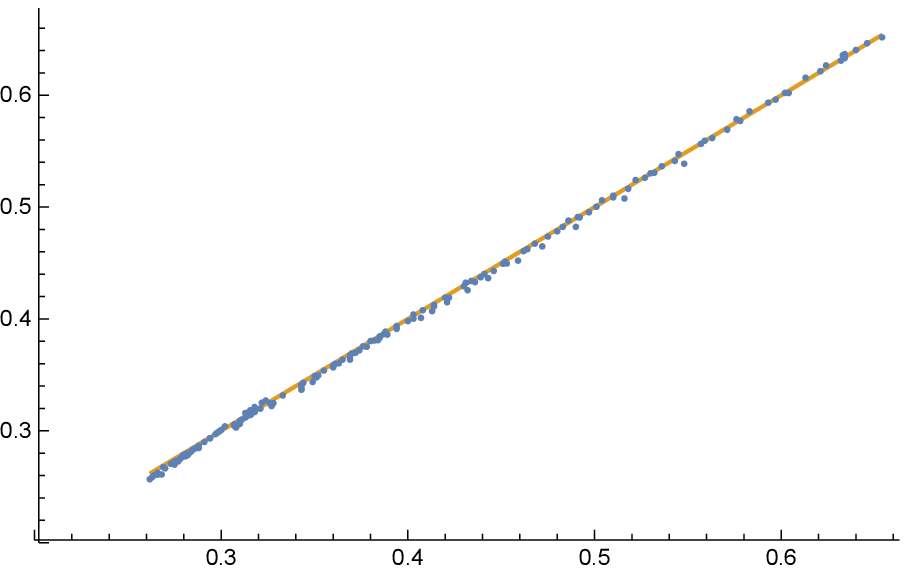}
\end{center}
  \caption{Calculated IV vs Closed Formula IV
}
 \label{fig:IVS}
\end{figure}

Using as input USD OIS, LIBOR rates and volatility of ATM Swaptions are taken at 3:04:41 pm on June 20, 2016 we generated 10,000 Monte-Carlo scenaros with semiannual steps and compared them with
closed form values. Results are shown on Fig.\ref{fig:IVS} where solid line are points Calculated IVs=Closed Formula IVs.

As we can see this formula works very well. Differences between closed form implied volatilities and Monte-Carlo results are about the size of stochastic noise.
It means that we can use this formula to fit all ATM swaption prices.

\section{Non Parametric Fit}

For every expiration we can calculate normal volatility for every tenor.
Most important is to interpolate the data in such a way to not increase volatility above needed value. The simplest way to do it is to set forward volatility to 0 if no
swaption data are available for selected expiration date. 
At any  expiration time  we need to estimate prices of all missed tenors. It is completed by using smooth splines.
Then we can calculate expected volatility for every time step assuming that first two of them has the same volatility.

It is not realistic assumption but it makes possible to identify possible arbitrage data points.
This procedure can be modified to more realistic one.

\begin{figure}[h]
\begin{minipage}{.5\textwidth}
\includegraphics[width=60mm]{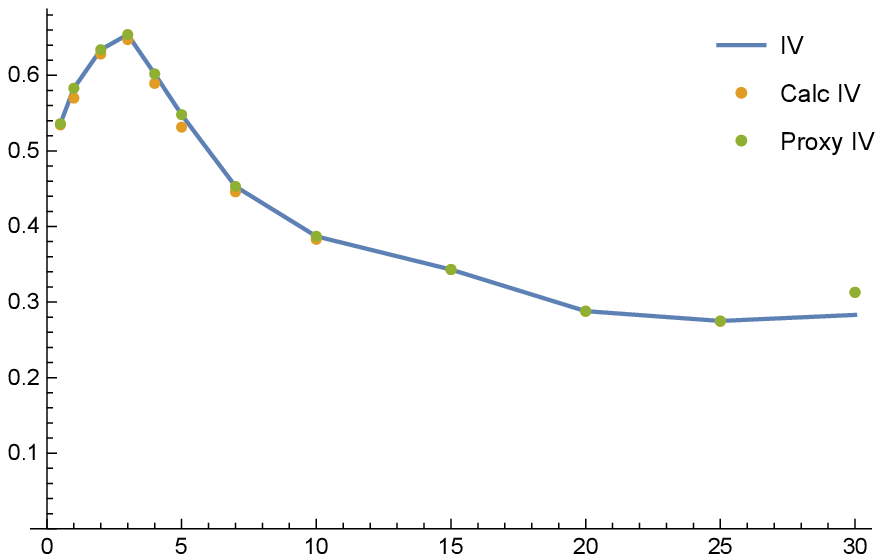}
  \caption{Tenor 1 IVs 
}
 \label{fig:ten1}
\end{minipage}
\begin{minipage}{.5\textwidth}
\includegraphics[width=60mm]{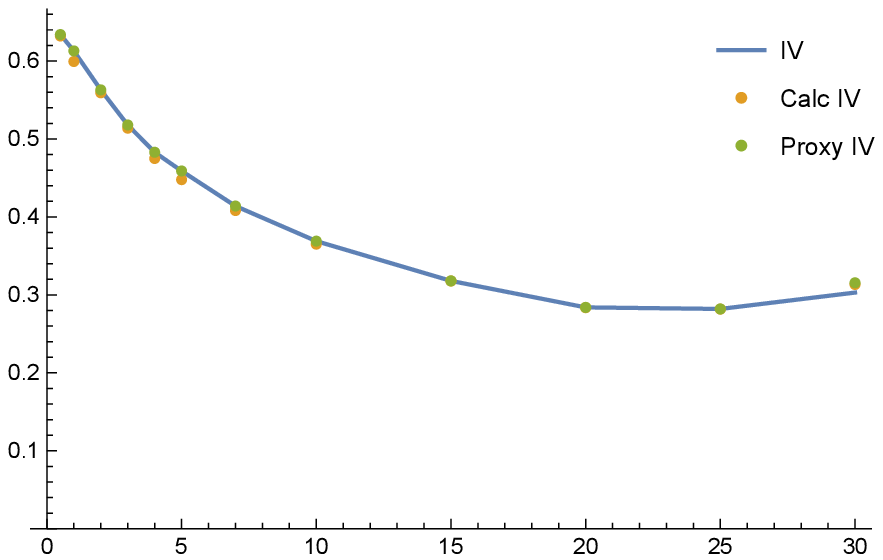}
  \caption{Tenor 5 IVs }
 \label{fig:ten5}
 \end{minipage}

\end{figure}

\begin{figure}[h]
\begin{minipage}{.5\textwidth}
\includegraphics[width=60mm]{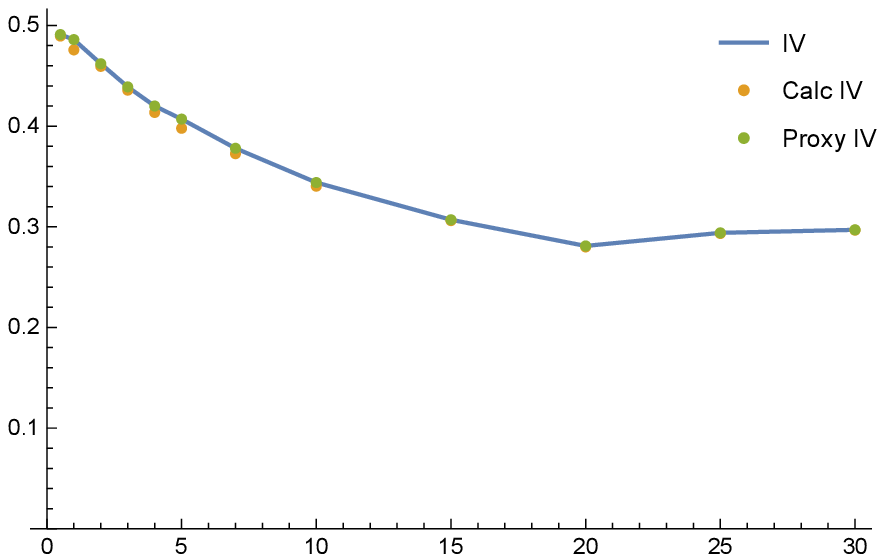}
  \caption{Tenor 10 IVs}
 \label{fig:ten10}
\end{minipage}
\begin{minipage}{.5\textwidth}
\includegraphics[width=60mm]{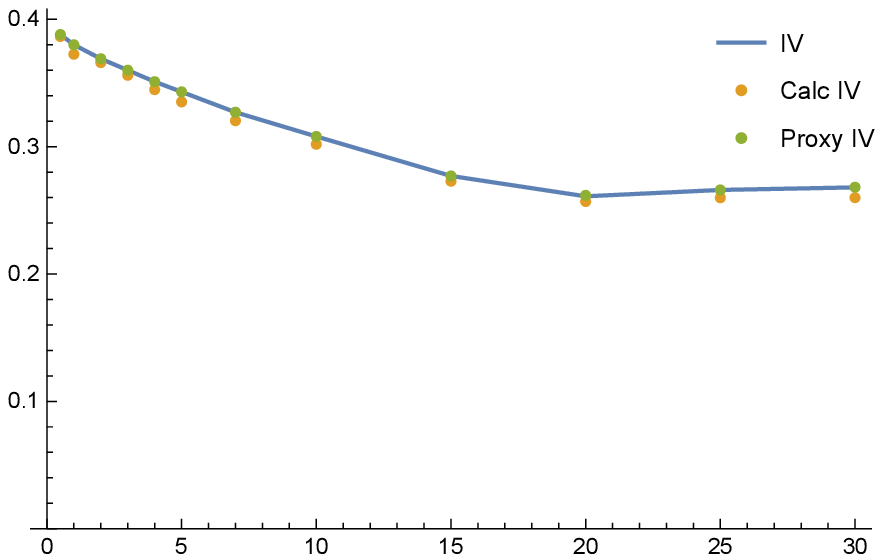}
  \caption{Tenor 30 IVs }
 \label{fig:ten30}
 \end{minipage}

\end{figure}

\begin{figure}[h]
\begin{center}
\includegraphics[width=60mm]{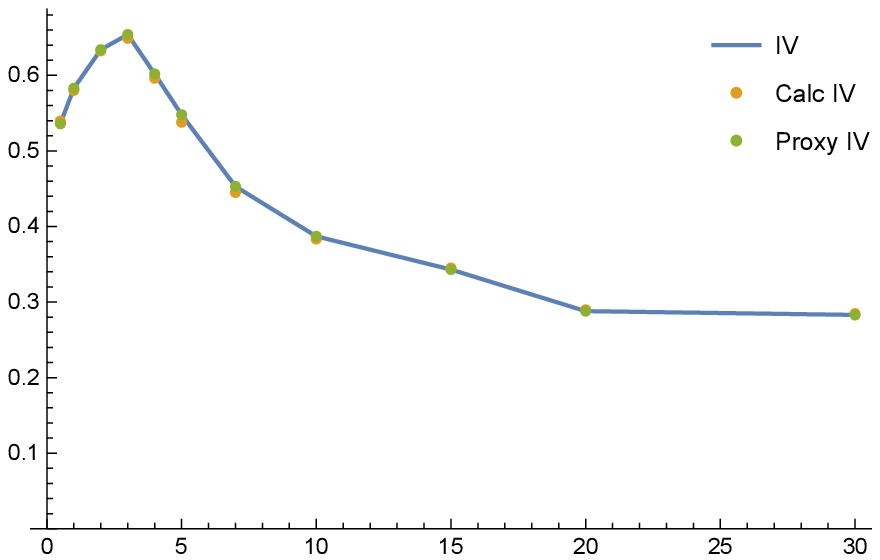}
  \caption{Tenor 1 IVs without 25 years to expiration}
   \label{fig:No25}
  \end{center}
\end{figure}

It works very good for all expirations and tenors. On
 Figs.\ref{fig:ten1}-\ref{fig:ten30}  calibration results are shown for Tenors 1, 5, 10 and 30 years.
 Input OIS, LIBOR rates and volatility of ATM Swaptions are taken at 3:04:41 pm on June 20, 2016.
 Note, that tenor 1 for 30 years to expiration can not be priced correctly. It means that either 30 years expiration is underpriced or 25 years expiration is overpriced.
 It was checked that this inconsistency can be removed excluding 25 expiration data point from calibration, see Fig.\ref{fig:No25}.


\section{Discussion}
Closed form formula for swaption prices in HJM model was obtained for single and dual curves.
 It was shown that closed form swaption formula for HJM model can be used to find a perfect fit of all ATM swaption prices.
If there is no way to fit the market prices it can be an indicator of arbitrage opportunity.

\end{document}